\documentclass[english]{article}
\usepackage{color}
\usepackage{amsthm}
\usepackage{amsmath}
\usepackage{graphicx}
\usepackage{amssymb}
\usepackage{esint}

\def\beq{\begin{equation}}
\def\eeq{\end{equation}}

\def\ben{\begin{eqnarray*}}
\def\een{\end{eqnarray*}}

\makeatletter

\numberwithin{equation}{section}
\numberwithin{figure}{section}

\theoremstyle{plain}
\newtheorem{thm}{Theorem}
  \theoremstyle{remark}
  \newtheorem{rem}[thm]{Remark}
\newcommand{\lyxaddress}[1]{
\par {\raggedright #1
\vspace{1.4em}
\noindent\par}
}

\makeatother

\usepackage{babel}

\begin{document}

\title{Scalar Quantum Field Theory on Fractals}

\author{Arnab Kar\footnote{arnabkar@pas.rochester.edu}  \ and S. G. Rajeev \footnote{rajeev@pas.rochester.edu} \footnote{Also at the Department of Mathematics}}

\maketitle

\begin{center}
Department of Physics and Astronomy\\ University of Rochester\\
Rochester NY 14627\\ USA
\end{center}

\begin{abstract}
We construct a family of measures for random fields based on the iterated
subdivision of simple geometric shapes (triangles, squares, tetrahedrons)
into a finite number of similar shapes. The intent is to construct
continuum limits of scale invariant scalar field theories, by imitating
Wiener's construction of the measure on the space of functions of
one variable. These are Gaussian measures, except for one example
of a non-Gaussian fixed point for the Ising model on a fractal. In the
continuum limits what we construct have correlation functions that vary
as a power of distance. In most cases this is a positive power (as
for the Wiener measure) but we also find a few examples with negative
exponent. In all cases the exponent is an irrational number, which
depends on the particular subdivision scheme used. This suggests that
the continuum limits corresponds to quantum field theories (random
fields) on spaces of fractional dimension.
\end{abstract}

\section{Introduction}

Wiener constructed a Gaussian measure on functions of one variable
by subdividing the interval and imposing the scale invariance of the
measure\cite{Ito on Wiener}. By subdividing the plane into triangles,
and space into tetrahedrons, we construct analogous measures for functions
of several variables. These are field theoretic analogues of the hierarchical
models\cite{Bethe Lattice} commonly studied in spin systems\cite{Hierarchical Model}.
They can also be viewed as the quantum analogues of the Laplace equation
on fractals studied by Strichartz\cite{Fractional Laplacian}. The
method we use is to seek fixed points of the {}``real space renormalization
group'', which has been in use in statistical physics for many years\cite{Kardar}. 

There are other approaches to constructing quantum field theories
in fractional dimensions \cite{Fractional QFT}, in part motivated
by the success of dimensional regularization. But our construction
appears to be more direct, being  based on a finite element method.
It does not appear to be possible to construct scale invariant Gaussian
measures of quantum field theories in integer space dimension greater
than one by our method of iterating finite integrals. We extend our
ideas of scale invariance to other systems like the Ising model, to find  a non trivial unstable fixed point for
a fractal of dimension greater than two. The critical exponents are
calculated in this case using ideas of statistical mechanics.

In the continuum limit what we construct is a quantum field theory
of fractional scaling dimension, with a correlation function that
varies as a power of distance; this power is often positive, but we
also find a few cases where it is negative. An example of a system
with growing correlation function would be that of Brownian motion.
The average distance two points undergoing Brownian motion increases
with time as $<\left[x(t_{1})-x(t_{2})\right]^{2}>\sim|t_{1}-t_{2}|$.

Physicists denote the Wiener measure by 

\[
\int{\cal D}\phi e^{-\frac{1}{2}\int\phi'^{2}dx}\]

Exactly what does this mean? There certainly is no Lebesgue measure
which can be called $\mathcal{D}\phi$ in the space of functions because
that space is infinite dimensional. Also, the functions $\phi$ are,
with probability one under the Wiener measure, not differentiable.
$\phi'$ is at best a distribution. So it does not have a sensible
product: $\phi'^{2}$ is by itself meaningless. 

Recall that the derivative $\frac{dy}{dx}$ in calculus is to be understood
as a limit constructed using epsilons and deltas of modern analysis,
and not as some infinitesimal divided by another. Similarly, the path
integral is to be understood as a limit of finite dimensional integrals.
The action being the integral of a local Lagrangian ( a function of
the field and its derivative) is not to be taken literally. Indeed,
a limit of a sequence of local measures (that depends only on nearest
neighbors of the triangulation) may not tend even formally to a local
action principle. 

In order to construct the scale invariant measures in higher dimensions,
the field variable could acquire an anomalous dimension. As a consequence,
the correlation functions have irrational numbers as their scaling
exponent. In the simplest examples,  $G(x_{1},x_{2})=<[\phi(x_1)-\phi(x_2)]^2>$ scales as $|x_{1}-x_{2}|^{0.611516}$
for the  subdivision of right angled triangled triangles and as $|x_{1}-x_{2}|^{0.329351}$
for tetrahedrons. These exponents are not universal as other subdivision
methods lead to other exponents.We suggest that this exponent of the correlation function of the Gaussian be used as a measure 
of the dimension of a fractal, analogous to   the Hausdorff dimension: it is more relevant for the behavior of randomness in quantum systems.

\section{Subdividing the Real Line}

Suppose that $\phi_{0}$ and $\phi_{1}$ are the boundary values of
the field on the interval $[x_{0},x_{1}]$ of length $a=x_{1}-x_{0}.$
Integrating out the field everywhere on the inside of the interval
induces some measure $h(\phi_{1},x_{1}|\phi_{0},x_{0})d\phi_{0}d\phi_{1}$
on the boundary values. The simplest possibility is a Gaussian that
is translation invariant in field values\[
h(\phi_{1},x_{1}|\phi_{0},x_{0})=C_{a}e^{-\frac{1}{2}\alpha_{a}(\phi_{1}-\phi_{0})^{2}}\]

for some $\alpha_{a}$ and $C_{a}$ which can depend on $|x_{1}-x_{0}|$.
If we subdivide the interval into two (for example, at its midpoint
$x_{01}=\frac{1}{2}(x_{1}+x_{0})$) and integrate over the field at
the new point, we should get back the original measure. \[
\int h(\phi_{1},x_{1}|\phi_{01},x_{01})h(\phi_{01},x_{01}|\phi_{0},x_{0})d\phi_{10}=h(\phi_{1},x_{1}|\phi_{0},x_{0})\]

This leads to the conditions

\ben
C_{\frac{a}{2}}^{2}\int\exp\left\{ -\frac{1}{2}\alpha_{\frac{a}{2}}\left[(\phi_{1}-\phi_{01})^{2}+(\phi_{01}-\phi_{0})^{2}\right]\right\} d\phi_{10}&=&C_{a}\exp\left\{ -\frac{1}{2}\alpha_{a}(\phi_{1}-\phi_{0})^{2}\right\} \nonumber \\
C_{a}&=&C_{\frac{a}{2}}^{2} \nonumber \\
\een

By completing the square,

\[
C_{\frac{a}{2}}^{2}\int\exp\left\{ -\frac{1}{2}\alpha_{\frac{a}{2}}\left[2\left(\phi_{01}-\frac{\phi_{0}+\phi_{1}}{2}\right)^{2}\right]\right\} d\phi_{10}\exp\left\{ -\frac{1}{2}\alpha_{\frac{a}{2}}\frac{(\phi_{1}-\phi_{0})^{2}}{2}\right\} =C_{a}\exp\left\{ -\frac{1}{2}\alpha_{a}(\phi_{1}-\phi_{0})^{2}\right\} \]

Thus,

\ben
\frac{1}{2}\alpha_{\frac{a}{2}}&=&\alpha_{a} \nonumber \\
C_{\frac{a}{2}}^{2}\sqrt{\frac{\pi}{\alpha_{\frac{a}{2}}}}&=&C_{a} \nonumber \\
\een

Solving

\ben
\alpha_{a}&=&\frac{\alpha}{a} \nonumber \\
C_{a}&=&\sqrt{\frac{\alpha}{2\pi a}} \nonumber \\
h(\phi_{1},x_{1}|\phi_{0},x_{0})&=&\frac{\sqrt{\alpha}}{\sqrt{2\pi(x_{1}-x_{0})}}e^{-\frac{\alpha}{2(x_{1}-x_{0})}(\phi_{1}-\phi_{0})^{2}} \nonumber \\
\een

By repeated subdivisions of the interval we would get the integral

\[
\int\prod_{i}d\phi_{i}e^{-\frac{\alpha}{2}\sum_{i}\frac{(\phi_{i+1}-\phi_{i})^{2}}{x_{i+1}-x_{i}}}\]
In the limit of small size of the interval, the exponent can be interpreted
as 

\[
\frac{\alpha}{2}\int\phi'^{2}dx\]

justifying the physicists' notation

\[
\int\prod_{i}d\phi_{i}e^{-\frac{\alpha}{2}\sum_{i}\frac{(\phi_{i+1}-\phi_{i})^{2}}{x_{i+1}-x_{i}}}\to\int{\cal D}\phi e^{-\frac{\alpha}{2}\int\phi'^{2}dx}.\]

But we reiterate that, this is not to be taken literally: $\lim_{x_{i+1}\to x_{i}}\frac{\phi_{i+1}-\phi_{i}}{x_{i+1}-x_{i}}$
almost never exists and so $\phi'^{2}$ is actually meaningless. The
quantity $\alpha$ may be related to mass divided by $\hbar$ in some
physical system or it may represent the inverse of the diffusion
coefficient for a different system.

A higher dimensional analogue of this process of subdivision would
involve triangulating space-time. We would subdivide each triangle
(or more generally simplex) to get smaller ones by introducing extra
vertices. Integrating out the field at the extra vertices should give
back the original measure.

\section{Subdividing Triangles}

Our idea is to approximate the functional integral of the quantum
field theory by an integration of the scalar values at the vertices
of a triangulation:

\[
\int\mathcal{D}\phi e^{-S[\phi]}A\left(\phi(x_{1}),\cdots\phi(x_{n})\right)\approx\int\prod_{i}d\phi_{i}\prod_{f}F_{a}\left(\phi(f)\right)A\left(\phi(x_{1}),\cdots\phi(x_{n})\right)\]

A triangulation is chosen with the points $x_{\alpha},\alpha=1,\cdots n$
in the argument of $A$ among the vertices. The measure of integration
is assumed to be local; i.e., can be expressed as a product of factors
$\ F_{f}\left(\phi_{0},\phi_{1},\phi_{2}\right)$ each depending only
on the field values at the vertices of a single face $f$ of the triangulation.
The basic idea is that if we subdivide a triangle by introducing new
vertices, and integrate over the field at these new vertices, we must
get back the contribution of the original triangle. 

What kind of triangle can be sub-divided into two equal triangles
which are each similar to the original one? A moment's reflection
shows that a right angled isosceles triangle has this property. Let
us work this out in two dimensions to get a feel for the method before
going to the much harder higher dimensional cases. However, it has
been shown that for higher dimensions, one can subdivide a $d$ simplex
in $n$ dimensions into $n^{d}$ similar simplexes. \cite{Subdivision of Simplex} 

\begin{figure}
\caption{\protect\includegraphics[scale=0.4]{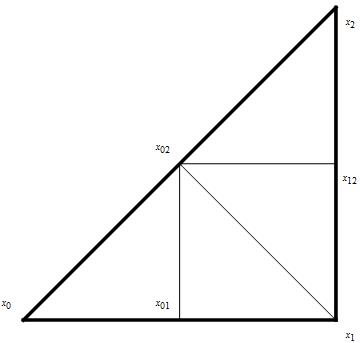}}
\label{Fig 1}
\end{figure}

We start with the triangle (Figure \ref{Fig 1}) with vertices $x_{0}=(0,0),\ x_{1}=(a,0),\ x_{2}=(a,a)$.
Let 

\[
F_{a}(\phi_{0},\phi_{1},\phi_{2})=C_{a}e^{-Q_{a}(\phi_{0},\phi_{1},\phi_{2})}\]

be the factor corresponding to such a triangle, for some normalization
constant $C_{a}$ and quadratic form $Q_{a}$. We assume that it is
invariant under translations in the field variables for simplicity.
It is likely that {}``massless'' scalar fields may have this property.
If the quadratic term was independent of cross terms between two vertices,
it may explain scalar fields in the {}``infinite mass'' limit. Also
the interchange of the vertices $1,\:2$ must be a symmetry. Thus
we have 

\[
Q_{a}(\phi_{0},\phi_{1},\phi_{2})=\alpha_{a}\left((\text{\ensuremath{\phi}}_{0}-\phi_{1})^{2}+(\text{\ensuremath{\phi}}_{2}-\text{\ensuremath{\phi}}_{1})^{2}\right)+\beta_{a}(\text{\ensuremath{\phi}}_{0}-\text{\ensuremath{\phi}}_{2})^{2}\]

for some constants $\alpha_{a},\:\beta_{a}.$ This can also be expressed
in matrix form as 

\[
Q_{a}(\phi_{0},\phi_{1},\phi_{2})=\phi^{T}\hat{Q}_{a}\phi,\quad\hat{Q}_{a}=\left(\begin{array}{ccc}
\alpha_{a}+\beta_{a} & -\alpha_{a} & -\beta_{a}\\
-\alpha_{a} & 2\alpha_{a} & -\alpha_{a}\\
-\beta_{a} & -\alpha_{a} & \alpha_{a}+\beta_{a}\end{array}\right)\]

If we choose the new internal vertices to be midpoints of the sides
of the original triangle $x_{0}x_{1}x_{2}$ \[
x_{ij}=\frac{1}{2}(x_{i}+x_{j}),\quad i,j=0,1,2\]

it divides the original triangle into the four equal triangles 
that are similar to the original. The new triangles $(x_{0}x_{01}x_{02,}\: x_{02}x_{01}x_{1},\: x_{1}x_{12}x_{02},\: x_{02}x_{12}x_{2})$ have orthogonal
edges which are half the length of the original triangle. 

The reproducibility condition obtained after integrating out the field
variables at the new vertices is

\[
e^{-Q_{2a}(\phi_{0},\phi_{1},\phi_{2})}=\int e^{-R_{a}(\phi_{0},\phi_{1},\phi_{2},\phi_{01},\phi_{02},\phi_{12})}d\phi_{01}d\phi_{02}d\phi_{02}\]

where, (up to a normalization constant)

\[
R_{a}(\phi_{0},\phi_{1},\phi_{2},\phi_{02})=Q_{a}(\phi_{0},\phi_{01},\phi_{02})+Q_{a}(\phi_{02},\phi_{01},\phi_{1})+Q_{a}(\phi_{1},\phi_{12},\phi_{02})+Q_{a}(\phi_{02},\phi_{12},\phi_{2})\]

This can be expressed in terms of a matrix form as 

\[
R_{a}(\phi_{0},\phi_{1},\phi_{2},\phi_{01},\phi_{02},\phi_{12})=\phi^{T}\hat{R}_{a}\phi\]

The result of integrating out the new variables $\phi_{01},\phi_{02},\phi_{12}$
is best understood by splitting the matrix $R$ into a block $L$
that acts on the original variables $\left(\phi_{0},\phi_{1},\phi_{2}\right)$
another $K$ acting on the new variables $\left(\phi_{01},\phi_{02},\phi_{12}\right)$
and a third matrix $J$ that mixes the two.

\[
\hat{R}_{a}=\left(\begin{array}{cc}
L & J\\
J^{T} & K\end{array}\right)\]

\[
L=\left(\begin{array}{ccc}
\alpha_{a}+\beta_{a} & 0 & 0\\
0 & 2\alpha_{a}+2\beta_{a} & 0\\
0 & 0 & \alpha_{a}+\beta_{a}\end{array}\right)\]

\[
K=\left(\begin{array}{ccc}
4\alpha_{a} & -2\alpha_{a} & 0\\
-2\alpha_{a} & 4\alpha_{a}+4\beta_{a} & -2\alpha_{a}\\
0 & -2\alpha_{a} & 4\alpha_{a}\end{array}\right)\]

\[
J=-\left(\begin{array}{ccc}
\alpha_{a} & \beta_{a} & 0\\
\alpha_{a} & 2\beta_{a} & \alpha_{a}\\
0 & \beta_{a} & \alpha_{a}\end{array}\right)\]

The exponent in the above Gaussian can also be written as 

\[
R_{a}(\phi_{0},\phi_{1},\phi_{2},\phi_{01},\phi_{02},\phi_{12})=\phi_{e}^{T}L\phi_{e}+2\phi_{e}^{T}J\phi_{i}+\phi_{i}^{T}K\phi_{i}\]

$\phi_{i}$ and $\phi_{e}$ denote the set of internal (new) and external
(original) vertices of the triangle. On completion of squares we get 

\[
R_{a}=\phi_{e}^{T}L\phi_{e}+(\phi_{i}^{T}+\phi_{e}^{T}J^{T}K^{-1})K(\phi_{i}+K^{-1}J\phi_{e})-\phi_{e}^{T}J^{T}K^{-1}J\phi_{e}\]

The result of the integral over the internal vertices then turns out
to be

\[
\hat{Q}_{2a}=L_{a}-J_{a}^{T}K_{a}^{-1}J_{a}\]

\[
=\left(\begin{array}{ccc}
\frac{1}{8}(5\alpha_{a}+6\beta_{a}) & \frac{1}{2}(-\alpha_{a}-\beta_{a}) & \frac{1}{8}(-\alpha_{a}-2\beta_{a})\\
\frac{1}{2}(-\alpha_{a}-\beta_{a}) & \alpha_{a}+\beta_{a} & \frac{1}{2}(-\alpha_{a}-\beta_{a})\\
\frac{1}{8}(-\alpha_{a}-2\beta_{a}) & \frac{1}{2}(-\alpha_{a}-\beta_{a}) & \frac{1}{8}(5\alpha_{a}+6\beta_{a})\end{array}\right)\]

\subsection{Fixed Points of the transformation}

The above result should be equal to the quadratic form

\[
\alpha_{2a}\left[\left(\phi_{0}-\phi_{1}\right)^{2}+\left(\phi_{1}-\phi_{2}\right)^{2}\right]+\beta_{2a}(\phi_{0}-\phi_{2})^{2}=\phi^{T}Q_{2a}\phi\]

On comparing its form we get 

\[
\alpha_{2a}=\frac{1}{2}(\alpha_{a}+\beta_{a}),\quad\beta_{2a}=\frac{1}{8}(\alpha_{a}+2\beta_{a})\]

If we define the ratio $x_{a}=\frac{\beta_{a}}{\alpha_{a}}$ we can
write these relations as 

\[
x_{2a}=f(x_{a}),\quad f(x)=\frac{1+2x}{4(1+x)}\]

\[
\alpha_{2a}=\lambda(x)\alpha_{a},\quad\lambda(x)=\frac{1}{2}(1+x)\]

We now seek a fixed point of the transformation $x_{a}\mapsto x_{2a}$
which is a quadratic equation with two roots

\ben
f(x)&=&x \nonumber \\
x&=&\frac{-1\pm\sqrt{5}}{4} \nonumber \\
\een

The solution with negative $x$ should be discarded as it does not
give an integrable Gaussian. The positive root corresponds to the
scaling relation

\[
\alpha_{2a}=\lambda\alpha_{a},\quad\lambda=\frac{1}{8}\left(3+\sqrt{5}\right)\]

Thus we set 

\[
x=\frac{\sqrt{5}-1}{4},\quad\lambda=\frac{1}{8}\left(3+\sqrt{5}\right)\]

If we make the ansatz $\alpha_{a}=a^{-\mu}\alpha_{1}$ we get 

\[
2^{-\mu}=\frac{3+\sqrt{5}}{8}\]

or \[
\mu=\frac{\log(\lambda^{-1})}{\log2}=\frac{\log\left(\frac{8}{3+\sqrt{5}}\right)}{\log2}\approx0.611516\]

In summary, we have a scale covariant solution to the reproducibility
condition 

\[
F_{a}(\phi_{0},\phi_{1},\phi_{2})=C_{a}\exp\left[-a^{-\mu}\alpha_{1}\left\{ (\text{\ensuremath{\phi}}_{0}-\phi_{1})^{2}+(\text{\ensuremath{\phi}}_{2}-\text{\ensuremath{\phi}}_{1})^{2}+\frac{\sqrt{5}-1}{4}(\text{\ensuremath{\phi}}_{0}-\text{\ensuremath{\phi}}_{2})^{2}\right\} \right]\]

We can also determine the normalization constant using the relation

\[
C_{a}^{4}\sqrt{\frac{\pi^{3}}{32(\alpha_{a}^{3}+2\alpha_{a}^{2}\beta_{a})}}=C_{2a}\]

but that is less interesting.

\subsection{Correlation function}

Choose two points $x_{1},x_{2}$ on the plane separated by some distance
$L=|x_{1}-x_{2}|.$ Draw an iso- right triangle with the line connecting
these points as a hypotnuse. We then bisect the right angle to get
two iso-right triangles, and subdivide them and so on. After a large
number of subdivisions we will get triangles of very small sides.
We then integrate out all vertices expect the original triangle to
calculate the correlation function:

\[
G(x_{1,},x_{2})=\frac{\int\prod_{i}d\phi_{i}\prod_{f}F_{a}\left(\phi(f)\right)\left[\phi(x_{1})-\phi(x_{2})\right]^{2}}{\prod_{i}d\phi_{i}\prod_{f}F_{a}\left(\phi(f)\right)}\]

The integrals can be evaluated using the reproducibility property
above. In the end we will end up with an integral over the two vertices
of the original large triangle whose side is $L$ :

\[
G(x_{1,},x_{2})=\frac{\int d\phi_{0}d\phi_{1}d\phi_{2}\left[\phi_{1}-\phi_{2}\right]^{2}\exp\left[-\frac{L^{-\mu}}{2}\alpha_{1}\left\{ \left((\text{\ensuremath{\phi}}_{0}-\phi_{1})^{2}+(\text{\ensuremath{\phi}}_{2}-\text{\ensuremath{\phi}}_{1})^{2}\right)+\frac{\sqrt{5}-1}{4}(\text{\ensuremath{\phi}}_{0}-\text{\ensuremath{\phi}}_{2})^{2})\right\} \right]}{\int d\phi_{1}d\phi_{2}\exp\left[-\frac{L^{-\mu}}{2}\alpha_{1}\left\{ \left((\text{\ensuremath{\phi}}_{0}-\phi_{1})^{2}+(\text{\ensuremath{\phi}}_{2}-\text{\ensuremath{\phi}}_{1})^{2}\right)+\frac{\sqrt{5}-1}{4}(\text{\ensuremath{\phi}}_{0}-\text{\ensuremath{\phi}}_{2})^{2})\right\} \right]}\quad\]

Thus, for some positive constant $G$, 

\[
G(x_{1},x_{2})=G|x_{1}-x_{2}|^{\mu}\sim|x_{1}-x_{2}|^{0.611516}\]

Note that the magnitude of the correlations increase with distance.
This is similar to the behavior of the average distance between two
points on a Brownian motion. On the other hand this is not the behavior
expected from a two dimensional quantum field theory with action $S=-\frac{1}{2}\int|\nabla\phi|^{2}d^{2}x=\frac{1}{2}(\phi,\Delta\phi)$.
The correlations behave logarithmically then

\[
\int\frac{d^{2}k}{(2\pi)^{2}}\frac{1}{|k|^{2}}e^{ik\cdot(x_{1}-x_{2})}\sim\log|x_{1}-x_{2}|\]

An interpretation is that it is a quantum field theory in a space
of fractal dimension in between one and two.

\section{Subdividing Tetrahedrons}

\begin{figure}
\caption{\protect\includegraphics[scale=0.4]{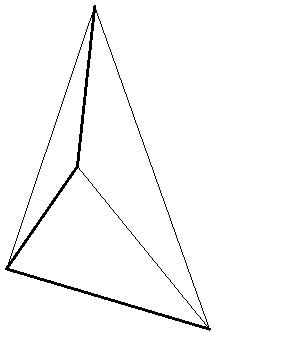}}
\label{Fig 2}
\end{figure}

It is not possible to subdivide a tetrahedron into two that are similar
to itself by a simple bisection: that would be equivalent to the ancient
problem of doubling the cube which is impossible by ruler-compass
methods (i.e., the cube root of $2$ cannot be found by such {}``Platonic''
constructions.) But it is possible to subdivide a tetrahedron into
eight tetrahedrons, each similar to the original. The trick is to
choose right isosceles tetrahedra, whose faces are all right angled
triangles, based on three mutually orthogonal edges of equal length\cite{Tetrahedron Subdivision}.
The figure \ref{Fig 2} shows an example of such a tetrahedron, the three mutually
orthogonal edges being shown in bold lines.

\begin{figure}
\caption{\protect\includegraphics[scale=0.5]{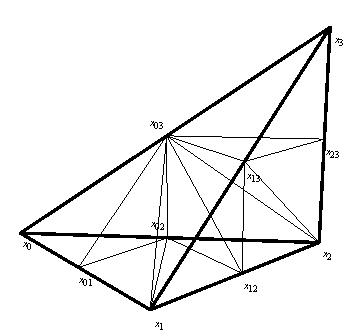}}
\label{Fig 3}
\end{figure}

To be more specific, the vertices 

\[
x_{0}=(0,0,0),\quad x_{1}=(a,0,0),\quad x_{2}=(a,a,0),\quad x_{3}=(a,a,a)\]

gives a right isosceles tetrahedron of smallest side $a$ (Figure \ref{Fig 3}). Every face
is a right angled triangle. There are three kinds of edges, of squared
lengths $a{}^{2},2a^{2},3a^{2}$. In fact, the squared length of the
edge $x_{ij}$ is just $|i-j|a^{2}$, for $i,j=0,1,2,3$. To each
such tetrahedron we associate a function $T_{a}(\phi_{0},\phi_{1},\phi_{2},\phi_{3})$
which represents the result of integrating out the field in its interior.

The internal vertices of the subdivision are the six points $x_{ij}=\frac{1}{2}(x_{i}+x_{j})$
for $0\leq i<j\leq3$. 

The reproducibility condition is the integral equation

\ben
T_{2a}(\phi_{0},\phi_{1},\phi_{2},\phi_{3})&=&\int T_{a}(\phi_{0},\phi_{01},\phi_{02},\phi_{03}) T_{a}(\phi_{1},\phi_{01},\phi_{02},\phi_{03}) \nonumber \\
&& T_{a}(\phi_{1},\phi_{12},\phi_{02},\phi_{03})T_{a}(\phi_{1},\phi_{12},\phi_{13},\phi_{03}) T_{a}(\phi_{2},\phi_{12},\phi_{02},\phi_{03}) \nonumber \\
&& T_{a}(\phi_{2},\phi_{12},\phi_{13},\phi_{03})T_{a}(\phi_{2},\phi_{23},\phi_{13},\phi_{03}) T_{a}(\phi_{3},\phi_{23},\phi_{13},\phi_{03}) \nonumber \\
&&  d\phi_{01}d\phi_{02}d\phi_{03}d\phi_{12}d\phi_{13}d\phi_{23}\label{eq:tetrlabel} \nonumber \\
\een

\begin{rem}
It is important to have the correct order of arguments in the $T_{a}$
as we are not dealing with a regular tetrahedron. We have verified
that with the orders given above, (different in detail from the reference
above) the distance matrices are exactly half the distance matrix
of the original tetrahedron.
\end{rem}
We assume a translation symmetry in field space (a global symmetry
we expect for massless fields)

\[
T_{a}(\phi_{0}+\phi,\phi_{1}+\phi,\phi_{2}+\phi,\phi_{3}+\phi)=T_{a}(\phi_{0},\phi_{1},\phi_{2},\phi_{3})\]

Also, assume a Gaussian ansatz for $T_{a}$. 

\ben
T_{a}(\phi_{0},\phi_{1},\phi_{2},\phi_{3})&=&e^{-Q_{a}(\phi_{0},\phi_{1},\phi_{2},\phi_{3})} \nonumber \\
Q_{a}(\phi_{0},\phi_{1},\phi_{2},\phi_{3})&=&\alpha\left((\text{\ensuremath{\phi_{0}}}-\text{\ensuremath{\phi_{1}}})^{2}+(\phi_{3}-\text{\ensuremath{\phi_{2}}})^{2}\right)+\beta\left((\text{\ensuremath{\phi_{0}}}-\text{\ensuremath{\phi_{2}}})^{2}+(\text{\ensuremath{\phi_{3}}}-\text{\ensuremath{\phi_{1}}})^{2}\right) \nonumber \\
&&+\gamma(\text{\ensuremath{\phi_{0}}}-\text{\ensuremath{\phi_{3}}})^{2}+\delta(\text{\ensuremath{\phi_{1}}}-\text{\ensuremath{\phi_{2}}})^{2} \nonumber \\
\een

This builds in the translation invariance as well as the reflection
symmetry $(0123)\rightleftarrows(3210)$ of the right isosceles tetrahedron.

The exponent of the integral (\ref{eq:tetrlabel}) is a quadratic
function of ten variables:

\ben
R_{a}(\phi_{0},\phi_{1},\phi_{2},\phi_{3},\phi_{01},\phi_{02},\phi_{03},\phi_{12},\phi_{13},\phi_{23})&=& Q_{a}(\phi_{0},\phi_{01},\phi_{02},\phi_{03}) +Q_{a}(\phi_{1},\phi_{01},\phi_{02},\phi_{03}) \nonumber \\
&&+Q_{a}(\phi_{1},\phi_{12},\phi_{02},\phi_{03}) +Q_{a}(\phi_{1},\phi_{12},\phi_{13},\phi_{03}) \nonumber \\
&&+Q_{a}(\phi_{2},\phi_{12},\phi_{02},\phi_{03})+Q_{a}(\phi_{2},\phi_{12},\phi_{13},\phi_{03}) \nonumber \\
&& +Q_{a}(\phi_{2},\phi_{23},\phi_{13},\phi_{03}) +Q_{a}(\phi_{3},\phi_{23},\phi_{13},\phi_{03}) \nonumber \\
\een

This is equal to $\frac{1}{2}\phi^{T}\hat{R}\phi$ for some $10\times10$
symmetric matrix $\hat{R}$, which can be calculated by evaluating
the second derivatives (Hessian), a straightforward procedure in Mathematica.
This matrix can be split into blocks

\[
\hat{R}=\left(\begin{array}{cc}
L & J\\
J^{T} & K\end{array}\right)\]

where $K$ is a matrix in the six dimensional subspace of the internal
variables $\left(\phi_{01},\cdots,\phi_{23}\right)$, $L$ is a matrix
in the four dimensional subspace of the external variables $\left(\phi_{0},\cdots,\phi_{3}\right)$
and $J$ is a $4\times6$ matrix that mixes them. The result of evaluating
the integral over the internal variables by completing the squares
is again a Gaussian with the Hessian matrix

\[
L-J^{T}K^{-1}J\]

\subsection{Fixed Points of the transformation}

Thus we can write the renormalization group transformation as 

\[
\hat{Q}_{2a}=L_{a}-J_{a}^{T}K_{a}^{-1}J_{a}\]

On comparing with the form of $Q_{a}$, the parameters $\alpha_{2a}\cdots\delta_{2a}$
will be determined as some homogeneous rational functions of $\alpha_{a}\cdots\delta_{a}.$The
ratios \[
\frac{\beta_{2a}}{\alpha_{2a}},\frac{\gamma_{2a}}{\alpha_{2a}},\frac{\delta_{2a}}{\alpha_{2a}}\]

depend only on the ratios $x=\frac{\beta_{a}}{\alpha_{a}},y=\frac{\gamma_{a}}{\alpha_{a}},z=\frac{\delta_{a}}{\alpha_{a}}$;
i.e. $\alpha_{a}$ and $\alpha_{2a}$ can be scaled out. Thus we get
a map of the projective space $P^{3}$ to itself. Each fixed point
$(x^{*},y^{*},z^{*})$ of this map is a Gaussian that is mapped into
itself under our transformation, except for an overall multiplication
of the quadratic form by $\frac{\alpha_{2a}}{\alpha_{a}}$ (the {}``eigenvalue''
$\lambda$) evaluated at $(x^{*},y^{*},z^{*})$. A sensible solution
must have a positive eigenvalue and also the parameters $(x^{*},y^{*},z^{*})$
must define a positive quadratic form $Q_{a}$. The equations for
the fixed points can be reduced to a polynomial for $x$

\[
-2+22x+132x^{2}+380x^{3}+699x^{4}+188x^{5}-1639x^{6}-2488x^{7}-948x^{8}+594x^{9}+608x^{10}+152x^{11}=0\]

The fixed point values for the other variables are determined in terms
of this root.

\ben
y&=&\frac{1}{107606064063}(-88214437948-533552485049x-1626406118465x^{2}-3995440895007x^{3} \nonumber \\
&&-4531514030431x^{4}+6599814153205x^{5}+18490138345312x^{6}+10764883382910x^{7} \nonumber \\
&&-3123153862430x^{8}-5175977687624x^{9}-1495324502248x^{10}) \nonumber \\
z&=&\frac{1}{35868688021}(-16477061906-533552485049x-1626406118465x^{2}-3995440895007x^{3} \nonumber \\
&&-4531514030431x^{4}+6599814153205x^{5}+18490138345312x^{6}+10764883382910x^{7} \nonumber \\
&&-3123153862430x^{8}-5175977687624x^{9}-1495324502248x^{10})\nonumber \\
\een

Of the eleven solutions, only five are real and among them exactly
one leads to a positive quadratic form $Q_{a}$

\[
x\approx0.699787174329122,\quad y\approx0.182495561519144,\quad z\approx2.547486684557433\]

The eigenvalue i.e. the ratio for this solution is $\frac{\alpha_{2a}}{\alpha_{a}}=\lambda\approx0.7958942$.

As for the right angled triangles, the correlation function can now
be calculated. We can choose any two vertices on the original tetrahedron
for this purpose and find the correlation between them. The choice
of vertices is not so important as the distance between them is very
large compared to the sides of the tetrahedrons obtained by subdivision.
The field variables in those vertices obtained by subdivision is integrated
out. The correlations will scale as $\lambda^{-1}$ when the distance
is doubled. That is, 

\[
G(x_{1},x_{2})\sim|x_{1}-x_{2}|^{\frac{\log\lambda}{\log2}}\sim|x_{1}-x_{2}|^{0.329351}\]

\section{Additional Examples}

\subsection{Gaussian Measures}

In both cases above we get a correlation function that increases with
distance. Are these exponents universal? Is it possible to find another
subdivision procedure that will produce a correlation that decreases
with distance? To answer these questions we worked out several additional
examples of subdivision. It turns out that each subdivision of the
plane (into squares, triangles of different shapes etc.) gives a different
exponent for the correlation function. Also, in most cases (except
two) we get correlation functions that grow with distance.

\subsubsection{Subdividing a Square}

\begin{figure}
\caption{\protect\includegraphics[scale=0.35]{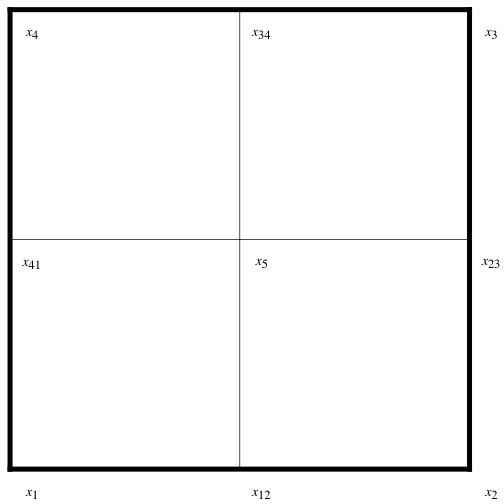}}
\label{Fig 4}
\end{figure}

We can also subdivide the plane into squares (Figure \ref{Fig 4}). To each square of side
$a$ we can associate a Gaussian weight with quadratic form 

\ben
Q_{a}(\phi_{1},\phi_{2},\phi_{3},\phi_{4})&=&\alpha_{a}\left[(\phi_{1}-\phi_{2})^{2}+(\phi_{2}-\phi_{3})^{2}+(\phi_{3}-\phi_{4})+(\phi_{4}-\phi_{1})^{2}\right] \nonumber \\
&&+\beta_{a}\left[(\phi_{1}-\phi_{3})^{2}+(\phi_{2}-\phi_{4})^{2}\right] \nonumber \\
\een

We subdivide each square by connecting the centroid $x_{5}$ to the
midpoints of the sides, $x_{12},x_{23},x_{34},x_{41}$. If we then
integrate out the five new vertices, we will get the transformations
as 

\[
\alpha_{2a}=\frac{1}{4}(2\alpha_{a}+\beta_{a}),\quad\beta_{2a}=\frac{2\alpha_{a}^{2}+4\alpha_{a}\beta_{a}+\beta_{a}^{2}}{8\alpha_{a}+4\beta_{a}}\]

\subparagraph*{Fixed Points and Correlation Function}

We get the condition for a fixed point of by setting $f(x_{a})=x_{2a}$
where $x_{a}=\frac{\beta_{a}}{\alpha_{a}}$:

\[
\frac{4\left(2+4x+x^{2}\right)}{(2+x)(8+4x)}=x\]

Of the three solutions, only $x=-1,\sqrt{3}-1$ lead to positive eigenvalues.

\[
\alpha_{2a}=\lambda\alpha_{a},\quad\lambda=\frac{1}{4},\mathrm{or}\ \frac{1}{4}\left(1+\sqrt{3}\right)\]

These two fixed points both give growing correlations

\[
G(x_{1}-x_{2})\sim|x_{1}-x_{2}|^{\mu},\quad\mu=2,\mathrm{or}\ 0.550016\]

\subsubsection{Apollonian Subdivision}

\begin{figure}
\caption{\protect\includegraphics[scale=0.4]{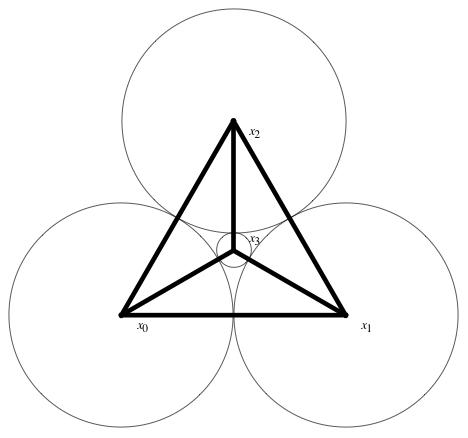}}
\label{Fig 5}
\end{figure}

Another example of a decreasing correlation function is the iterated
Apollonian subdivision of a triangle: we join the vertices to an interior
point, thereby subdividing the triangle into three triangles (Figure \ref{Fig 5}). Another
interpretation is in terms of circles: we can regard the plane as
being subdivided into three mutually tangent circles centered at the
vertices of a triangle. Then we can insert a new circle in the interstitial
region that is tangent to the other three circle. This process can
then be repeated. Every tangency of a pair of circles corresponds
to an edge connecting their centers. 

The scaling law of the correlation function can be computed just as
before by computing the Gaussian integral over the new vertex $x_{3}$

\[
e^{-\alpha_{\sqrt{3}a}\left[(\phi_{0}-\phi_{1})^{2}+(\phi_{0}-\phi_{2})^{2}+(\phi_{1}-\phi_{2})^{2}\right]}\propto\]

\[
\int e^{-\alpha_{a}\left[(\phi_{0}-\phi_{1})^{2}+(\phi_{0}-\phi_{2})^{2}+(\phi_{1}-\phi_{2})^{2}+2(\phi_{0}-\phi_{3})^{2}+2(\phi_{1}-\phi_{3})^{2}+2(\phi_{2}-\phi_{3})^{2}\right]}d\phi_{3}\]

Each side emanating from vertex $x_{3}$ is shared by two triangles,
which explains the factor of two in those terms. The proportionality
is given as the scaling is not done rigorously here. We say that one
third of the vertices are integrated out roughly after each step and
we get $\sqrt{3}$ as the scale factor in two dimensions.

\subparagraph*{Fixed Points and Correlation Function}

This leads to the transformation

\[
\alpha_{\sqrt{3}a}=\frac{5}{3}\alpha_{a}\]

and a correlation function with exponent 

\[
G(x_{1},x_{2})\sim|x_{1}-x_{2}|^{\mu},\quad\mu=\frac{\log\frac{3}{5}}{\log\sqrt{3}}\approx-0.929947\]

Finally we see an example of a correlation decreasing with distance.
This may be interpreted as a fractal with dimension greater than two.

\subsubsection{Dividing a triangle into six similar triangles}

\begin{figure}
\caption{\protect\includegraphics[scale=0.4]{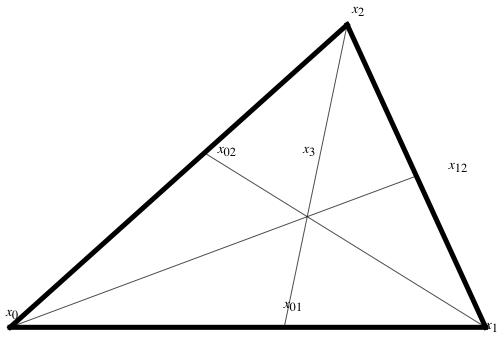}}
\label{Fig 6}
\end{figure}

Another example of a decreasing correlation function is the division
of an equilateral triangle divided up into six equilateral triangles (Figure \ref{Fig 6}).
Topologically this is the same as the barycentric subdivision, or
the bisection of the angles of the triangles\cite{Bisectors Triangles}:
we add a new point at the centroid (on the in-center) and connect it
to the original vertices and midpoints of the sides. But as long as
the six smaller triangles occupy the same plane as the original one,
it is not possible for them all to be equilateral. But we can assign
equal lengths to all the edges as long as we do not insist that the
graph has an isometric embedding in Euclidean space. By iterating
the subdivision process, we will get a fractal which makes sense as
a metric space, but not as a sub-manifold of Euclidean space.

To each of the equilateral triangle we associate the Gaussian weight
\[
Q_{a}(\phi_{0},\phi_{1},\phi_{2})=\alpha_{a}\left[(\phi_{0}-\phi_{1})^{2}+(\phi_{1}-\phi_{2})^{2}+(\phi_{2}-\phi_{0})^{2}\right]\]

The field variables at the vertices $\left(x_{01},x_{02},x_{12}\right)$
and the center of the original triangle $(x_{3})$ are integrated
over to give a bigger equilateral triangle with the vertices
$\left(x_{0},x_{1},x_{2}\right).$ A side of the newly formed triangle
is bigger than the original triangle's side by a factor of $\sqrt{6}.$ 

\[
e^{-Q_{\sqrt{6}a}(\phi_{0},\phi_{1},\phi_{2})}=\int e^{-R_{a}(\phi_{0},\phi_{1},\phi_{2},\phi_{3},\phi_{01},\phi_{02},\phi_{12})}d\phi_{3}d\phi_{01}d\phi_{12}d\phi_{12}
\]

with 

\ben
R_{a}(\phi_{0},\phi_{1},\phi_{2},\phi_{3},\phi_{01},\phi_{02},\phi_{12})&=&Q_{a}(\phi_{0},\phi_{01},\phi_{3})+Q_{a}(\phi_{0},\phi_{3},\phi_{02})+Q_{a}(\phi_{1},\phi_{3},\phi_{01}) \nonumber \\
&&+Q_{a}(\phi_{1},\phi_{12},\phi_{3})+Q_{a}(\phi_{2},\phi_{3},\phi_{12})+Q_{a}(\phi_{2},\phi_{02},\phi_{3}) \nonumber \\
\een

\subparagraph*{Fixed Points and Correlation Function}

On computing the Gaussian integral and comparing it with the Gaussian
form expected we get the following transformation

\[
\alpha_{\sqrt{6}a}=\frac{5}{4}\alpha_{a}\]

The new quadratic form for the triangle formed will be 

\[
Q_{\sqrt{6}a}(\phi_{0},\phi_{1},\phi_{2})=\frac{5}{4}\alpha_{a}\left[(\phi_{1}-\phi_{0})^{2}+(\phi_{0}-\phi_{2})^{2}+(\phi_{1}-\phi_{2})^{2}\right]
\]

Thus the correlation function scales with exponent as 

\[
G(x_{1}-x_{2})\sim|x_{1}-x_{2}|^{\mu},\quad\mu=\frac{\log\frac{4}{5}}{\log\sqrt{6}}\approx-0.249078\]

Thus it is possible to get a correlation function that decreases with
distance as well.

\subsection{Ising Models}

\subsubsection{The One dimensional Ising Model }

We carry out the idea of sub division of the real line once more.
However we are now interested in the Ising model now and our field
values $(\sigma_{i})$ at the end points of the of the interval can
be $\pm1.$ Instead of integrals we end up doing a discrete sum. Let
the function associated with an interval be 

\ben
F_{a}(\sigma_{0},\sigma_{1})&=&e^{Q_{a}(\sigma_{0},\sigma_{1})} \nonumber \\
Q_{a}(\sigma_{0},\sigma_{1})&=&\alpha_{a}(\sigma_{0}\sigma_{1})+\gamma_{a} \nonumber \\
\een

There will not be any higher order terms of $\sigma_{i}$ since $\sigma_{i}^{2}=1$
and $\sigma_{i}\sigma_{j}=i\epsilon_{ijk}\sigma_{k}$. We divide the
interval into two halves now and the new variable is summed over. 

\ben
\sum_{\sigma_{01}}e^{Q_{a}(\sigma_{0},\sigma_{01})}e^{Q_{a}(\sigma_{01},\sigma_{1})}&=&\sum_{\sigma_{01}}e^{\alpha_{a}\sigma_{01}(\sigma_{0}+\sigma_{1})+2\gamma_{a}} \nonumber \\
&=&\left[e^{\alpha_{a}(\sigma_{0}+\sigma_{1})}+e^{-\alpha_{a}(\sigma_{0}+\sigma_{1})}\right]e^{2\gamma_{a}} \nonumber \\
\een

\subparagraph*{Fixed Points and Correlation Function}

On comparing with the function associated with the whole interval
\[
F_{2a}(\sigma_{0},\sigma_{1})=e^{\alpha_{2a}(\sigma_{0}\sigma_{1})+\gamma_{2a}}\]

we get the following pair of equations considering all the combinations
formed by the pair $\left(\sigma_{0},\sigma_{1}\right)$

\ben
2e^{2\gamma_{a}}&=&e^{\gamma_{2a}-\alpha_{2a}} \nonumber \\
e^{2\gamma_{a}}\left(e^{2\alpha_{a}}+e^{-2\alpha_{a}}\right)&=&e^{\gamma_{2a}+\alpha_{2a}} \nonumber \\
\een

The transformation obtained from these equations is 

\[
e^{2\alpha_{2a}}=\frac{e^{2\alpha_{a}}+e^{-2\alpha_{a}}}{2}\]

The trivial fixed point $\alpha_{a}=0$ is obtained by setting $\alpha_{2a}=\alpha_{a}.$
The normalization constant can be determined from the relation between
$\gamma_{a}$ and $\gamma_{2a}$.

\subsubsection{Ising Model on the Fractal Subdivision of a Right Angled Triangle}

\begin{figure}
\caption{\protect\includegraphics[scale=0.35]{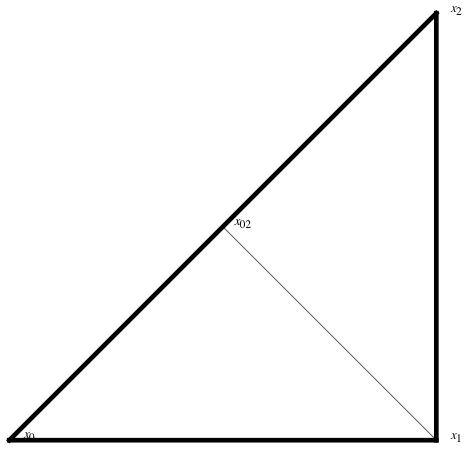}}
\label{Fig 7}
\end{figure}

We carry out the idea of sub division for a triangle now (Figure \ref{Fig 7}). Let the
function associated with the triangle be 

\ben
F_{a}(\sigma_{0},\sigma_{1},\sigma_{2})&=&e^{Q_{a}(\sigma_{0},\sigma_{1},\sigma_{2})} \nonumber \\
Q_{a}(\sigma_{0},\sigma_{1},\sigma_{2})&=&\alpha_{a}(\sigma_{0}\sigma_{1}+\sigma_{1}\sigma_{2})+\beta_{a}(\sigma_{0}\sigma_{2})+\gamma_{a} \nonumber \\
\een

We now divide the original triangle in two new similar triangles.
The new vertex is summed over and the total contribution from the
two triangles is computed.

\ben
\sum_{\sigma_{02}}e^{Q_{a}(\sigma_{1},\sigma_{02},\sigma_{0})}e^{Q_{a}(\sigma_{2},\sigma_{02},\sigma_{1})}&=&\sum_{\sigma_{02}}e^{\alpha_{a}\sigma_{02}(\sigma_{0}+\sigma_{2}+2\sigma_{1})+\beta_{a}\sigma_{1}(\sigma_{0}+\sigma_{2})+2\gamma_{a}} \nonumber \\
&=&\left[e^{\alpha_{a}(\sigma_{0}+\sigma_{2}+2\sigma_{1})}+e^{-\alpha_{a}(\sigma_{0}+\sigma_{2}+2\sigma_{1})}\right]e^{\beta_{a}\sigma_{1}(\sigma_{0}+\sigma_{2})+2\gamma_{a}} \nonumber \\
\een

\subparagraph*{Fixed Points and Correlation Function}

On comparing it with the form of the energy associated with the bigger
triangle 

\[
F_{\sqrt{2}a}(\sigma_{0},\sigma_{1},\sigma_{2})=e^{\alpha_{\sqrt{2}a}\sigma_{1}(\sigma_{0}+\sigma_{2})+\beta_{\sqrt{2}a}(\sigma_{0}\sigma_{2})+\gamma_{\sqrt{2}a}}\]

we get the following pairs of equations

\ben
e^{2(\gamma_{a}+\beta_{a})}\left(e^{4\alpha_{a}}+e^{-4\alpha_{a}}\right)&=&e^{2\alpha_{\sqrt{2}a}+\beta_{\sqrt{2}a}+\gamma_{\sqrt{2}a}} \nonumber \\
e^{2\gamma_{a}}\left(e^{2\alpha_{a}}+e^{-2\alpha_{a}}\right)&=&e^{-\beta_{\sqrt{2}a}+\gamma_{\sqrt{2}a}} \nonumber \\
2e^{2(\gamma_{a}-\beta_{a})}&=&e^{-2\alpha_{\sqrt{2}a}+\beta_{\sqrt{2}a}+\gamma_{\sqrt{2}a}} \nonumber \\
\een

In order to obtain these equations we look into all the tuples formed
by assigning values to $\left(\sigma_{0},\sigma_{1},\sigma_{2}\right).$
Out of the set of eight equations we get only three independent equations.
The transformations obtained from these equations are 

\ben
e^{4\alpha_{\sqrt{2}a}}&=&\frac{e^{4\beta_{a}}\left(e^{4\alpha_{a}}+e^{-4\alpha_{a}}\right)}{2} \nonumber \\
e^{4\beta_{\sqrt{2}\alpha}}&=&\frac{2\left(e^{4\alpha_{a}}+e^{-4\alpha_{a}}\right)}{\left(e^{2\alpha_{a}}+e^{-2\alpha_{a}}\right)^{2}} \nonumber \\
\een

The trivial fixed points $\alpha_{a}=0,\:\beta_{a}=0$ are obtained
by setting $\alpha_{\sqrt{2}a}=\alpha_{a},\:\beta_{\sqrt{2}a}=\beta_{a}$.
$\gamma_{a}$ and $\gamma_{\sqrt{2}a}$ determine the normalization
factor.

\subsubsection{Ising Model on sixfold subdivision of the triangle}

After having found trivial fixed points $\alpha_{a}=0$ till now we
take up the example of the sixfold subdivision of the triangle to
see if it has a non trivial fixed point. And indeed it has one. To
each of the equilateral triangles we associate a function as shown
below. 
\ben
F_{a}(\sigma_{0},\sigma_{1},\sigma_{2})&=&e^{Q_{a}(\sigma_{0},\sigma_{1},\sigma_{2})} \nonumber \\
Q_{a}(\sigma_{0},\sigma_{1},\sigma_{2})&=&\alpha_{a}(\sigma_{0}\sigma_{1}+\sigma_{1}\sigma_{2}+\sigma_{2}\sigma_{0})+\gamma_{a} \nonumber \\
\een

\lyxaddress{A sum is now carried over the vertices 
$(\sigma_{01},\sigma_{02},\sigma_{12},\sigma_{3})$ to form a bigger
equilateral triangle with the vertices  $(\sigma_{0},\sigma_{1},\sigma_{2})$. }

\ben
&&\sum_{\sigma_{01}}\sum_{\sigma_{02}}\sum_{\sigma_{12}}\sum_{\sigma_{3}}e^{Q_{a}(\sigma_{1},\sigma_{12},\sigma_{3})}e^{Q_{a}(\sigma_{2},\sigma_{12},\sigma_{3})}e^{Q_{a}(\sigma_{2},\sigma_{02},\sigma_{3})}e^{Q_{a}(\sigma_{0},\sigma_{02},\sigma_{3})}e^{Q_{a}(\sigma_{0},\sigma_{01},\sigma_{3})}e^{Q_{a}(\sigma_{1},\sigma_{01},\sigma_{3})} \nonumber \\
&=&e^{-6\alpha_{a}+6\gamma_{a}}(2+e^{-2(-2+\sigma_{2}+\sigma_{0})\alpha_{a}}+e^{2(2+\sigma_{2}+\sigma_{0})\alpha_{a}}+e^{-2(-2+\sigma_{2}+\sigma_{1})\alpha_{a}}+e^{2(2+\sigma_{2}+\sigma_{1})\alpha_{a}} \nonumber \\
&&+e^{-2(-2+\sigma_{0}+\sigma_{1})\alpha_{a}}+e^{2(2+\sigma_{0}+\sigma_{1})\alpha_{a}}+e^{-4(-3+\sigma_{2}+\sigma_{0}+\sigma_{1})\alpha_{a}}+e^{4(3+\sigma_{2}+\sigma_{0}+\sigma_{1})\alpha_{a}} \nonumber \\
&&+e^{-2(-4+2\sigma_{2}+\sigma_{0}+\sigma_{1})\alpha_{a}}+e^{2(4+2\sigma_{2}+\sigma_{0}+\sigma_{1})\alpha_{a}}+e^{-2(-4+\sigma_{2}+2\sigma_{0}+\sigma_{1})\alpha_{a}}+e^{2(4+\sigma_{2}+2\sigma_{0}+\sigma_{1})\alpha_{a}} \nonumber \\
&&+e^{-2(-4+\sigma_{2}+\sigma_{0}+2\sigma_{1})\alpha_{a}}+e^{2(4+\sigma_{2}+\sigma_{0}+2\sigma_{1})\alpha_{a}}) \nonumber \\
\een

\subparagraph*{Fixed Points and Correlation Function}

The function associated with the bigger triangle should be of the
form 
\[
F_{\sqrt{6}a}(\sigma_{0},\sigma_{1},\sigma_{2})=e^{\alpha_{\sqrt{6}a}(\sigma_{1}\sigma_{0}+\sigma_{0}\sigma_{2}+\sigma_{2}\sigma_{1})+\gamma_{\sqrt{6}a}}
\]

We get the following pairs of equations on comparing the two results.

\ben
e^{-\alpha_{\sqrt{6}a}+\gamma_{\sqrt{6}a}}&=&e^{-6(\alpha_{a}-\gamma_{a})}\left(3+6e^{4\alpha_{a}}+4e^{8\alpha_{a}}+2e^{12\alpha_{a}}+e^{16\alpha_{a}}\right) \nonumber \\
e^{3\alpha_{\sqrt{6}a}+\gamma_{\sqrt{6}a}}&=&e^{-6(\alpha_{a}-\gamma_{a})}\left(9+3e^{8\alpha_{a}}+3e^{16\alpha_{a}}+e^{24\alpha_{a}}\right) \nonumber \\
\een

Thus the renormalization group transformation on $\alpha$ is the
iteration of the function 

\[
f(\alpha)=\frac{-1}{4}\text{log}\left[\frac{3+6e^{4\alpha_{a}}+4e^{8\alpha_{a}}+2e^{12\alpha_{a}}+e^{16\alpha_{a}}}{9+3e^{8\alpha_{a}}+3e^{16\alpha_{a}}+e^{24\alpha_{a}}}\right]\]

There is a trivial fixed point for the equation $\alpha_{a}=f(\alpha_{a})$
at $\alpha_{a}=0$ and a non trivial fixed point at $\alpha_{a*}=0.159579$.
The normalization constants are determined by $\gamma_{a}$ and $\gamma_{\sqrt{6}a}.$ 

To find  the critical exponents we linearize the recursion relation
at the fixed point\cite{Kardar}. 

\[
f'(\alpha_{a*})=1.93411\]

This shows that the fixed point is unstable. The scaling factor for
this transformation is $\sqrt{6}.$ Using these numbers we can calculate
the thermal exponent $y_{t}\simeq\frac{\log(1.93411)}{\log(\sqrt{6})}=0.736312$.
This leads to the critical exponents for the correlation length and
specific heat to be $\nu=\frac{1}{y_{t}}=1.35812,\:\alpha=2(1-\nu)=-0.716239$. 

$ $

\[
\]

\section{Acknowledgement}

It is a pleasure to thank Sreedhar B. Dutta, Tamar Friedman, Govind
Krishnaswami, Yannick Meurice, Fred Moolekamp, Ambar Sengupta and S.
Shankaranarayanan for discussions. This work was supported in part
by a grant from the US Department of Energy under contract DE-FG02-91ER40685.

\end{document}